\documentclass[conference]{IEEEtran}
\IEEEoverridecommandlockouts
\usepackage{cite}
\usepackage{amsmath,amssymb,amsfonts}
\usepackage{algorithmic}
\usepackage{graphicx}
\usepackage{tabularx}
\usepackage{array}
\usepackage{textcomp}
\usepackage{xcolor}
\usepackage{url}
\usepackage{dblfloatfix}
\usepackage{booktabs} 
\usepackage{multirow}
\usepackage{makecell}
\usepackage{graphicx}
\usepackage{rotating} 
\usepackage{subcaption}
\usepackage{siunitx}
\def\BibTeX{{\rm B\kern-.05em{\sc i\kern-.025em b}\kern-.08em
    T\kern-.1667em\lower.7ex\hbox{E}\kern-.125emX}}
\begin{document}

\title{ClusterBench: A Framework for Cluster-Wide Continuous Benchmarking and Regression Testing}

\makeatletter
\newcommand{\linebreakand}{%
  \end{@IEEEauthorhalign}
  \hfill\mbox{}\par
  \mbox{}\hfill\begin{@IEEEauthorhalign}
}
\makeatother

\author{
  \IEEEauthorblockN{Aditya Ujeniya}
  \IEEEauthorblockA{\textit{Erlangen National HPC Center}\\
    \textit{Friedrich-Alexander-University}\\
    Erlangen, Germany \\
    0009-0000-1670-4131}
  \and
  \IEEEauthorblockN{Jan Eitzinger}
  \IEEEauthorblockA{\textit{Erlangen National HPC Center}\\
    \textit{Friedrich-Alexander-University}\\
    Erlangen, Germany \\
    0009-0000-3350-3841}
  \and
  \IEEEauthorblockN{Thomas Gruber}
  \IEEEauthorblockA{\textit{Erlangen National HPC Center}\\
    \textit{Friedrich-Alexander-University}\\
    Erlangen, Germany \\
    0000-0001-5560-6964}
  \linebreakand % <--- Forces new row and centers remaining authors
  \IEEEauthorblockN{Georg Hager}
  \IEEEauthorblockA{\textit{Erlangen National HPC Center}\\
    \textit{Friedrich-Alexander-University}\\
    Erlangen, Germany \\
    0000-0002-8723-2781}
  \and
  \IEEEauthorblockN{Gerhard Wellein}
  \IEEEauthorblockA{\textit{Erlangen National HPC Center}\\
    \textit{Friedrich-Alexander-University}\\
    Erlangen, Germany \\
    0000-0001-7371-3026}
}

\maketitle

\begin{abstract}
Data centers require tooling that validates the stability and reliability of an entire installation rather than of individual nodes. Such validation is needed when a cluster is accepted into service and at regular intervals for the rest of its lifetime. This requires that identical benchmarks be dispatched, in a single submission, to every node of the cluster, which in turn requires a cluster-aware scheduling framework. This paper presents ClusterBench, a framework for cluster-wide continuous benchmarking with cluster-aware scheduling. A benchmark collection ships with ClusterBench to target each component: CPU, GPU, memory, interconnect, and I/O disk. Running this suite through ClusterBench tests each component within a node and collects the statistical variation across cross-specimen components in the whole cluster. Because measurements are repeated throughout the cluster's lifetime, ClusterBench collects data across both space and time. Comparing measurements against earlier runs allows detection of performance regressions introduced by software changes, such as OS kernel updates or new library versions. The repeated measurements also constitute a comprehensive dataset for research on hardware variability.

Across component-targeted benchmarks on the NHR@FAU clusters Helma, Alex, and Fritz, variation within a single component stays within 1\%. Depending on the component and benchmark type, performance variation across specimens reaches up to 5\%, despite nodes that are identical by specification. To better understand this variation, performance is further correlated with collected metrics such as power draw, frequency, and temperature. This correlation depends on how the data is viewed: combining space (i.e., specimens) and time presents a different picture than considering either dimension alone, and the relationship between temperature and performance also differs between air-cooled and liquid-cooled nodes. 

\end{abstract}


\section{Introduction}

Data centers validate a cluster at multiple stages in its life: at acceptance, when the administrators must show that the delivered system meets its specification, and then at regular intervals for the rest of its service life. The latter is necessary because neither the hardware nor the software stack stays fixed. Components age and eventually fail, firmware settings change, kernels are patched, and compilers, MPI implementations, and numerical libraries are replaced. Any of these can change the performance of end user applications.

In both cases, validation must cover the entire cluster rather than any single node. The relevant question is not whether one node meets its expected performance, but whether all nodes do, and how widely the measured values are distributed. This matters for multi-node jobs, because the slowest process determines the runtime of the whole job. Validation must therefore measure every node, and do so in a way that makes the results comparable.

Existing benchmarking frameworks make the second part difficult. A test is specified by the resources it requests, and the choice of nodes is left to the scheduler --- appropriate for the question these frameworks were built to answer: how a benchmark performs at a given scale. Some frameworks support cluster-wide benchmarking through single-node, cluster-aware scheduling, but not with the granularity needed to test every component within a node individually. Moreover, no existing framework provides collecting metrics such as power consumption, frequency, and temperature out of the box, which would be valuable for correlating with the performance of each individual component. Section~\ref{sec:related} examines these limitations in detail and motivates a framework designed specifically to address them.

This paper presents ClusterBench\footnote{\url{https://github.com/ClusterCockpit/cc-clusterbench}}, a framework for cluster-wide continuous benchmarking and regression testing. A test is defined once, for a cluster or one of its partitions. ClusterBench then submits the jobs, tracks their completion, and stores each result together with the node, the time, and the software environment. Coverage can be exhaustive: every node can be measured on its own, node pairs can be measured to characterize the interconnect, and on nodes with several accelerators, each accelerator can be exercised in parallel. Tests can also run on a schedule, so results accumulate along two axes. Across nodes, the spread of a metric over nominally identical nodes reveals the state of the installation and highlights outliers. Over time, comparison with earlier runs of the same test exposes regressions from software changes, as well as the slow hardware degradation that can precede a component failure.

Our benchmark collection targets each hardware component separately: CPU, GPU, memory, interconnects, and disks. A component-level benchmark points to one component, while an application-level result mixes them all. ClusterBench also wraps each benchmark in a metric collector, so power, clock frequency, and temperature are recorded over the same interval as the benchmark. These metrics can help explain part of the spread observed along both axes --- spatial and temporal. The consequence is that exact reproducibility of time-based metrics such as Flops/s and Bytes/s cannot be guaranteed even on a correctly functioning node; reproducibility is a meaningful claim only when a margin of statistical deviation is allowed.

Our contributions are as follows.

\begin{itemize}
  \item identify cluster-aware scheduling as the capability required to validate a whole cluster installation, and define it to cover not only one benchmark per node, but also node pairs for multi-node benchmarks and individual accelerators within a node, all derived from a single definition of an input test file.
  \item report a component-based variability study of production clusters, separating within-component from cross-specimen variation for CPUs and GPUs.
  \item correlate the observed performance variation with power, clock frequency, and temperature to see how strongly they correlate.
\end{itemize}

Section~\ref{sec:related} reviews benchmarking frameworks, benchmarking suites, and earlier studies of hardware variability. 

Section~\ref{sec:architecture} describes in detail the design and features of ClusterBench, Section~\ref{sec:benchsuite} the benchmark suite and the measurement method, and Section~\ref{sec:results} the results. Section~\ref{sec:limitations} further discusses ClusterBench's current limitations and what it is not meant for.

\section{Related Work}
\label{sec:related}

\subsection{Benchmarking and regression frameworks}

Running a benchmark reproducibly requires several steps: the environment must be prepared, the code built, a job submitted, and the figures of merit extracted. Several centres have combined these steps into frameworks. JUBE\footnote{\url{https://github.com/FZJ-JSC/jube}} describes a benchmark in XML or YAML. It executes the benchmark as a sequence of dependent steps. Each step is instantiated once for every combination of a parameter space. Results are written as tables or into a SQLite database~\cite{jube_parco}. ReFrame\footnote{\url{https://github.com/reframe-hpc/reframe}} defines a test as a Python class. Each test passes through a fixed pipeline. The pipeline builds the code, submits the job, polls its state, and checks sanity and performance. Each session is stored in a SQLite database~\cite{reframe_hust2019}. Pavilion2\footnote{\url{https://github.com/hpc/pavilion2}} describes tests in YAML, and most parts of the system can be extended through plugins~\cite{pavilion2,chicoma_cug2021}. Similar frameworks include buildtest, the OLCF Test Harness, BenchPRO, and Ramble~\cite{buildtest,olcf_harness,benchpro,ramble}. AutoBench~\cite{autobench} targets HPC testbeds. It defines knobs in layers for partition, node, OS, software stack, and benchmark. Every combination of these knobs becomes one benchmark instance. Jobs are generated from templates and started from CI. The extracted figures of merit are then joined with telemetry from the DCDB\footnote{\url{https://gitlab.lrz.de/dcdb/dcdb}} monitoring system. Benchpark\footnote{\url{https://github.com/LLNL/benchpark}} describes a benchmark as a Spack package that builds it and a Ramble application that runs it. These specifications are held in a shared repository. An experiment defined at one centre can therefore be repeated at another~\cite{benchpark}.

Some of these frameworks can cover a whole machine, and they do so in a similar way. ReFrame can distribute a test to all nodes in a given state. Pavilion2 can divide the system into chunks and run a test on each chunk. AutoBench can name partitions, nodes, and components as knobs and emit one job for each. The node is a coordinate in the configuration space rather than a participant in one experiment. The authors state this limitation themselves and list large jobs across a cluster as future work. Benchpark differs for another reason. A benchmark exists there as a recipe for building the code. Its specifications also serve comparisons between centres rather than between the nodes of one machine. ReFrame is limited further with the \texttt{--distribute} mode, because it distributes only single-node tests, and only for some scheduler backends. An interconnect benchmark is therefore excluded. Covering the interconnect of a cluster requires forming node pairs from the node list; separating single-node from multi-node benchmarks within one job allocation; or dispatching one instance per accelerator inside a node, with systematic collection of the respective output for all these component-level testing cases. None of these frameworks offer this out of the box, so users must configure and adapt the framework themselves, and component-level testing may not fit within a single definition file.

This difference also changes how output is stored. In the common case, one job runs one benchmark on one node with one configuration, so writing the output once per job is enough to tell results apart. JUBE, ReFrame and AutoBench therefore organize their output by benchmark. ClusterBench cannot rely on this, because a single allocation may carry measurements for many nodes, for node pairs, and for individual accelerators. The job alone no longer identifies a result, so the placement must become part of the output path. Without it, the files of a run cannot be traced back to the units that produced them. ClusterBench therefore derives a fixed directory layout from the configuration and the placement. The extraction step uses the same layout to find an output file together with the metric samples that belong to it.

The frameworks also differ in the metrics they collect during a run. ReFrame and Pavilion2 record only the figure of merit reported by the benchmark. Values such as power draw, frequency, and temperature are therefore not available. Such values sometimes explain a large part of the difference between runs. AutoBench does associate them with its results. It obtains them by querying DCDB. This monitoring database runs continuously on the target system and is maintained independently of the benchmarking platform. The metrics are therefore joined to a result after the run has finished. The approach also requires that monitoring is already deployed for the machine. 
 
The form in which results are kept is a further consideration. JUBE, ReFrame, and Pavilion2 use row-major stores such as SQLite. AutoBench writes the extracted metrics as CSV into measurement repositories. Both forms serve as a record of what has been run, and both are suitable for retrieving a single result. They are less suitable as a basis for aggregation. Measurements accumulate across many nodes, components, and repetitions over the lifetime of a system, and their analysis then benefits from a columnar organization of the data.

All the existing frameworks are feature-rich and have mature ecosystem include compilation phase for benchmarks before executing. ClusterBench purposefully excludes it and we explain more about the reasons to exclude this in the Limitations and Future Work Section~\ref{sec:compilation_phase}.

\subsection{Continuous benchmarking}
Another related work concerns repeating benchmarks over time. The motivation is that continuous integration checks correctness, while performance is typically left unchecked~\cite{exacb}. exaCB, developed at JSC for JUPITER, is the closest work to ours in this group~\cite{exacb}: it integrates benchmarking into CI/CD pipelines and delegates execution to JUBE framework, with its components connected through a versioned JSON protocol. Results that conform to this protocol are kept, so they can be read as time series, showing whether the system has regressed after a change to the software stack. AutoBench serves a similar purpose: daily runs started from CI are used to monitor the health of hardware components~\cite{autobench}. Further related work includes Thicket, which supports the analysis of performance data collected across many runs~\cite{thicket}, and the reproducible-benchmarking principles proposed by Koskela et al.~\cite{koskela2023}.

These frameworks are driven by changes in software. A pipeline runs when the code or the software stack changes, and the measurement shows whether that change affected performance. Here, the code is the object under test, and the machine is the fixture. Our approach is different. We hold the binary fixed. Instead, the environment around it changes: the operating system, drivers, firmware, and the state of the hardware over the machine's lifetime. So a run starts on a weekly schedule, or after an event such as acceptance testing or the end of maintenance, rather than after a code commit. This changes what we measure. A single value per benchmark records that something has changed, but not where. A set of values covering all nodes at one point in time separates a shift of the whole machine from the deviation of an individual node, and this distribution can be the quantity of interest.

\subsection{Benchmarking Suites}
Several benchmark suites already exist. The NAS Parallel Benchmarks are a small set of programs derived from computational fluid dynamics codes, made up of five kernels and three pseudo-applications, with problem sizes fixed in advance as classes~\cite{bailey1991nas}. The HPC Challenge suite adds seven tests, among them HPL, STREAM, DGEMM, PTRANS, RandomAccess, FFT, and a latency and bandwidth test~\cite{luszczek2005hpcc,dongarra2013hpcc}. The JUPITER Benchmark Suite combines 16 application benchmarks with 7 synthetic ones and was built for the procurement of an exascale system~\cite{herten2024jupiter}. CORAL-2 serves the same purpose, and asks vendors to tune its codes to show the value of the hardware they offer.\footnote{\url{https://asc.llnl.gov/coral-2-benchmarks}}

Some of these suites share individual benchmarks with ours, and the JUPITER suite is planned for the acceptance procedure of the new system~\cite{herten2024jupiter}. Acceptance testing, however, happens once. A health check happens throughout the lifetime of the cluster, daily or weekly, and must run on a machine that is otherwise in production. For this purpose an application benchmark is a poor probe. It touches many components at the same time, so a change in its result does not say which part of the node has degraded. A component-level benchmark gives a single performance number for a single piece of hardware, and that number can be compared against the value measured when the machine was new.

In the proposed benchmark collection, each benchmark is tied to one component of a node: CPU, GPU, main memory, GPU memory, network interconnects, and local disk. Where a vendor-tuned build exists, that build is used, so the component runs at peak performance, good enough to stress the desired component. A portable reference (vanilla or non-optimized) code  reaches only part of the peak performance and does not stress the component enough.

\subsection{Variation between identical hardware}

Measurements have repeatedly contradicted the assumption that nodes of the same specification perform identically. Rountree et al.\ reported power differences between processors of the same type~\cite{rountree2012}, and Inadomi et al.\ measured about two thousand CPU and DRAM modules and attributed the observed variation to manufacturing variability~\cite{inadomi2015}. The same effect has been described under Turbo Boost~\cite{turbo_ics2016}, across processor generations~\cite{marathe2017}, in a production system at NERSC~\cite{cori_power}, and across several hundred GPUs~\cite{gpu_variation_icpp2022}. Process variation changes the achievable clock frequency and the leakage current of individual dies, and frequency scaling turns this into a difference in performance. 

So far, all these studies have been conducted only with application benchmarks. A variability study of component-level benchmarks and the CPU/GPU related metrics has not been done yet.

\section{ClusterBench architecture and features}
\label{sec:architecture}

ClusterBench is designed to operate in daemon mode. The daemon mode requires several background workers — a cron submitter, a job tracker, and a node list populator — and is additionally necessary for the submission of benchmarks. Go was selected as the implementation language due to its strong concurrency support, as native goroutines allow workers to be spawned and managed with minimal code complexity. Furthermore, the framework benefits from Go’s robust built-in tooling, comprehensive standard library, and straightforward deployment. Benchmark definitions are accepted only in JSON, which is parsed by the standard libraries without any external dependence. Its grammar is also unambiguous, avoiding both the implicit type coercion of YAML and any dependence on indentation or whitespace. With reference to Figure~\ref{fig:architecture}, the architecture of ClusterBench is explained in the following sections.

\begin{figure}[tbp]
    \centering
    \includegraphics[width=0.49\textwidth]{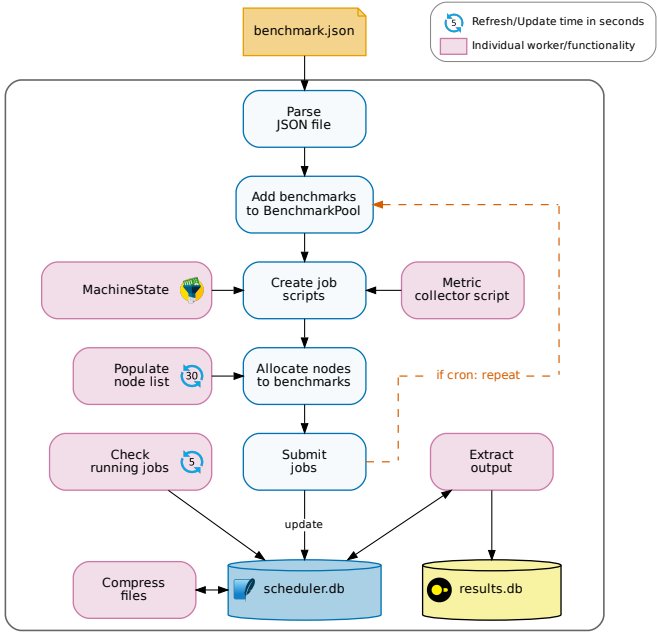}
    \caption{Architecture of ClusterBench framework}
    \label{fig:architecture}
\end{figure}

\subsection{General pipeline}

ClusterBench takes a simple definition file as input. It parses this file and loads the applicable benchmarks into the BenchmarkPool, discarding any invalid benchmark definitions. At startup, ClusterBench also populates a list of the idle and allocated nodes across the different partitions; this list is then used to assign nodes to the jobs that will be submitted. The next step is to generate structured job files and submit them to the assigned nodes. Into each job a call to MachineState\footnote{\url{https://github.com/RRZE-HPC/MachineState}}\cite{machinestate} is injected, which records the state of a node before the benchmarks begin. Capturing this state makes it possible to detect later changes in software, modules or kernel versions, any of which may account for the deviations observed during regression testing. Once node assignment is complete, all jobs are submitted simultaneously and recorded in a book-keeping database, from which other workers can retrieve the information they need.

\subsection{Databases}
ClusterBench ships with two persistent database objects. Embedded databases are preferable to external database services, as they avoid introducing an additional runtime dependency on the framework.

\subsubsection{Book-keeping database}
A persistent SQLite database records all submitted jobs and their state. Because SQLite is row-oriented, it is well suited for storing per-job records such as the partition and nodes a benchmark was submitted to and whether it completed or failed.

\subsubsection{Analytics database}
A persistent DuckDB database stores the output produced by the benchmarks. Because DuckDB is column-oriented, it supports the fast aggregate queries needed to build responsive dashboards. As metrics can be collected per benchmark, the output extractor records both the benchmark result itself and the associated metrics, such as frequency, temperature and power draw. With all of these values held in a single database, users can visualize them in the dashboarding tool of their choice; the persistent database object can be used directly as a data source in Grafana, Elasticsearch, or similar tools.

\subsection{Metric-collector script}

ClusterBench ships with a Perl-based metric collector. When metric collection is enabled for a benchmark in its definition file, the benchmark is wrapped in a collector that runs alongside it and writes the hardware metrics of the interval to a separate file. For CPUs, temperature and clock frequency are read from the sysfs interface, and power is read from the kernel powercap interface, which requires RAPL support on the processor. For GPUs, the metrics are queried from \texttt{nvidia-smi} or \texttt{amd-smi}. Because the collector is bound to the benchmark, the metrics cover exactly the interval that the benchmark was running for. Metric collection therefore depends on the node configuration. Power readings require that the powercap files are readable by the job user, which is not the default on recent kernels. Where a source is unavailable, the affected metric is omitted and the remaining ones are still recorded.

\subsection{Background workers}

\subsubsection{Node-list populator}
This worker runs at startup and, by default, populates lists of the nodes that are idle or allocated in each available partition. The lists are refreshed every 30 seconds, which is particularly useful for periodically scheduled (cron) benchmarks. The worker can also be configured to track other node states, such as drained, down, reserved or under maintenance, although only the idle and allocated states are considered by default.

\subsubsection{Running-job checker}
ClusterBench offers two ways to check whether the benchmarks have completed or not: normal update mode, which runs only once during the startup of the framework, or a separate worker that checks and updates the benchmark every specified interval. A separate worker inspects the running jobs every 5 seconds (configurable) and updates their state in the database as either completed or failed. 
%The user is also allowed to configure other values than the default interval of 5 seconds.

\subsubsection{Output extractor}
This functionality selects all completed and failed jobs from the book-keeping database \texttt{scheduler.db} and searches their output for the values of interest. Extraction can be performed in two ways. The first is to specify a regular expression, from which the first numerical value in the matched pattern is taken. The second is to supply a path to a user-provided shell script, which receives inputs such as the output file name, the cluster name and the partition name, processes the output file and returns a single numerical value. The extracted values are then written to the analytics database \texttt{results.db}. Moreover, if the metric measurement was selected in benchmark definition file and if the metric measurement file exists for the respective benchmark, then it will also average all the metrics over time, and store an averaged value in the database. ClusterBench also offers to extract LIKWID\footnote{\url{https://github.com/RRZE-HPC/likwid}}\cite{likwid} measurements over sysfs measurements. If the LIKWID measurements are present in the output file, then the user can configure to extract them instead of averaging the measurements from sysfs metrics.

\subsubsection{File compressor}
This functionality compresses the various output files once the extraction phase has completed; extraction is therefore a prerequisite for compression. Because the framework supports cluster-wide and component-level benchmarking and can additionally record per-benchmark metrics, the volume of output it produces grows rapidly. Compressing the output files on a per-benchmark basis reduces both the disk space consumed and the number of files retained.

\section{Benchmark collection}
\label{sec:benchsuite}

The benchmark collection\footnote{\url{https://github.com/RRZE-HPC/benchmark-collection}} is a curated collection of node-level benchmarks assembled for continuous regression testing of production HPC systems. Its design follows a component-oriented principle: every benchmark is associated with exactly one hardware component within a single compute node, and it is configured such that it places the highest possible load or stress on that component. The relevant components are the CPU, the GPU, main memory, GPU device memory, the communication fabric, and the local disk. Since each measurement is intended to approach the attainable hardware limit rather than to represent a particular application, the resulting figures can be compared against vendor specifications and, more importantly, against earlier runs on the same hardware. The suite is deployed through ClusterBench, which schedules the benchmarks across nodes and collects the results.

The individual benchmarks are described in the following.

\paragraph{HPL}
The High Performance Linpack (HPL)\footnote{\url{https://www.netlib.org/benchmark/hpl}} benchmark solves a dense system of linear equations by LU decomposition and is the de-facto standard for reporting floating-point throughput of HPC systems. We use it to characterize the compute performance of the CPU, and rely on vendor-optimized builds because these reach peak floating-point performance at a comparatively small matrix size. A vanilla HPL build, by contrast, must be configured with a matrix that occupies roughly 80\% of main memory, which results in run times well in excess of 30 minutes and is therefore impractical for health checking the nodes in production clusters.

\paragraph{DGEMM and SGEMM}
On the GPU we use dense matrix--matrix multiplication from the vendor BLAS libraries as the compute-bound counterpart of HPL, since it exhibits high arithmetic intensity and drives the floating-point units close to their limit. Depending on the ratio of FP64 to FP32 hardware units, either DGEMM (double precision) or SGEMM (single precision) is the appropriate choice. HPL itself is not suitable in this role: the VRAM capacity of a GPU is far smaller than the main memory of a node, so only modest matrix sizes can be allocated, and the resulting run completes within a few minutes. Such a short run is sufficient to report a peak figure but leaves the device under load for too little time to serve as a stress test. A GEMM kernel, in contrast, can be repeated for a reasonable
duration at a constant load level.

\paragraph{TheBandwidthBenchmark}
This benchmark\footnote{\url{https://github.com/RRZE-HPC/TheBandwidthBenchmark}} comprises a set of streaming kernels that differ in their data traffic and in their ratio of floating-point operations to transferred bytes. It is executed on both the CPU and the GPU and yields the saturated bandwidth of main memory and of GPU device memory, respectively. Recent server-grade NVIDIA GPU architectures such as Blackwell and Rubin provide an increasingly wide memory interface, and saturating such an interface requires a sufficient number of bytes in flight. TheBandwidthBenchmark therefore offers vectorized load and store variants of its kernels, which are necessary to reach the attainable bandwidth on such devices.

\paragraph{OSU Micro-Benchmarks}
The OSU Micro-Benchmarks\footnote{\url{https://mvapich.cse.ohio-state.edu/benchmarks}} cover a comprehensive range of MPI communication patterns and also support GPU device buffers. They are straightforward to deploy and include dedicated tests for GPU-aware communication. We use the multi-rank CPU-only benchmarks to characterize and stress test the inter-node communication fabric. For the intra-node GPU interconnect, such as NVLink and NVSwitch, we use the all-to-all communication pattern, since it stresses all links simultaneously and thus reveals the aggregated latency of the interconnect.

\paragraph{fio}
The Flexible I/O Tester (fio)\footnote{\url{https://github.com/axboe/fio}} generates synthetic I/O workloads from a job description that specifies the access pattern, block size, number of concurrent jobs, and queue depth, and reports bandwidth, I/O operations per second, and latency distributions. We apply it to the node-local disk with direct I/O enabled, so that the page cache is bypassed and the measurement characterizes the storage device itself. Since read and write paths differ in cost, the reported bandwidth strongly depends on the mix of the two. 
Currently, we use a ratio of 20\% reads and 80\% writes,
%is something we are experimenting, 
because the lifespan of an SSD is governed by the volume of data written before its performance degrades, and a write-dominated workload is therefore the more demanding and the more relevant case.

\medskip

Table~\ref{tab:benchsuite} summarizes the benchmarks together with the component they target and the metric they report. All benchmarks are integrated as Git submodules and are built through a common Makefile, so that the whole suite can be obtained and compiled with a single recursive clone. This keeps the individual benchmarks at their upstream sources and makes the exact version used for a given measurement reproducible.

\begin{table}[!t]
  \renewcommand{\arraystretch}{1.15}
  \caption{Benchmarks in the collection, the node component they are aimed at, and the reported metric.}
  \label{tab:benchsuite}
  \centering
  \footnotesize
  \setlength{\tabcolsep}{4pt}
  \begin{tabularx}{\columnwidth}{@{}
      >{\raggedright\arraybackslash}p{0.32\columnwidth}
      >{\raggedright\arraybackslash}p{0.24\columnwidth}
      >{\raggedright\arraybackslash}X@{}}
    \hline
    Benchmark & Target component & Metric \\
    \hline
    HPL\textsuperscript{*}            & CPU                  & Floating-point performance \\
    DGEMM/SGEMM\textsuperscript{*}    & GPU                  & Floating-point performance \\
    TheBandwidthBenchmark             & CPU and GPU memory   & Memory bandwidth \\
    OSU Micro-Benchmarks         & Communication fabric & Bandwidth and latency \\
    fio                               & Local disk           & I/O bandwidth, IOPS, latency \\
    \hline
  \end{tabularx}
 
  \vspace{2pt}
  {\footnotesize \textsuperscript{*}Vendor-optimized implementation.}
\end{table}

\section{Cluster-wide statistical variability of CPU and GPU}
\label{sec:results}

\subsection{Testbed and methodology}

%We present the specs of our production clusters on which the study was performed. 
Table \ref{tab:nhr_cluster_specs} summarizes the details of the CPUs, GPUs, and interconnects within each cluster and partition used for the tests presented in this study.

\begin{table*}[t]
\centering
\caption{Hardware specifications across NHR@FAU clusters and different partitions}
\label{tab:nhr_cluster_specs}
\footnotesize
\setlength{\tabcolsep}{2.5pt} % Precise padding for 7-column fit

\begin{tabular}{l cc cc cc}
\toprule
\textbf{Cluster} & \multicolumn{2}{c}{\textbf{Alex}} & \multicolumn{2}{c}{\textbf{Helma}} & \multicolumn{2}{c}{\textbf{Fritz}} \\
\cmidrule(lr){2-3} \cmidrule(lr){4-5} \cmidrule(lr){6-7}

\textbf{Partition} & \texttt{a100} & \texttt{a40/a100} & \texttt{h100} & \texttt{cpu} & \texttt{singlenode} & \texttt{spr2tb} \\
\cmidrule(lr){2-2} \cmidrule(lr){3-3} \cmidrule(lr){4-4} \cmidrule(lr){5-5} \cmidrule(lr){6-6} \cmidrule(lr){7-7}

\textbf{Target} & \textbf{GPU} & \textbf{Host CPU} & \textbf{GPU} & \textbf{CPU} & \textbf{CPU} & \textbf{CPU} \\
\midrule

\textbf{Chip Name} & \makecell{NVIDIA\\A100} & \makecell{AMD EPYC\\7713} & \makecell{NVIDIA\\H100} & \makecell{AMD EPYC\\9965} & \makecell{Intel Xeon\\8360Y} & \makecell{Intel Xeon\\8470} \\
\textbf{Microarch.} & Ampere & Zen 3 & Hopper & Zen 5c & Ice Lake & Sapphire Rapids \\
\textbf{SMs / Cores} & 108 SMs & 64 Cores & 132 SMs & 192 Cores & 36 Cores & 52 Cores \\
\textbf{Per Node} & 8 GPUs & 2 Sockets & 4 GPUs & 2 Sockets & 2 Sockets & 2 Sockets \\
\textbf{\# Nodes} & 20 / 18 & 82 & 96 & 312 & 992 & 48 \\
\textbf{Device TDP} & 400\,W & 225\,W & 700\,W & 500\,W & 250\,W & 350\,W \\
\textbf{Base Freq.} & 1.09\,GHz & 2.00\,GHz & 1.6\,GHz & 2.25\,GHz & 2.40\,GHz & 2.00\,GHz \\
\textbf{Boost Freq.} & 1.41\,GHz & 3.68\,GHz & 1.98\,GHz & 3.70\,GHz & 3.50\,GHz & 3.80\,GHz \\
\textbf{Peak FP64 FLOPS} & 19.4\,T & 2.05\,T & 33.45\,T & 6.90\,T & 2.76\,T & 3.33\,T \\
\textbf{Mem. Capacity} & \makecell{40\,GB /\\80\,GB} & \makecell{1\,TB /\\2\,TB} & 94\,GB & 768\,GB & 256\,GB & 2\,TB \\
\textbf{Mem. BW} & 1.5\,TB/s / 2\,TB/s & 204.8\,GB/s & 2.41\,TB/s & 460.8\,GB/s & 204.8\,GB/s & 307.2\,GB/s \\
\textbf{Mem. Tech} & HBM2 & DDR4 & HBM2e & DDR5 & DDR4 & DDR5 \\
\textbf{Mem. Clock} & 1215\,MHz / 1593\,MHz & 3200\,MT/s &  1593\,MHz & 5600\,MT/s & 3200\,MT/s & 4800\,MT/s \\
\textbf{Interconnect} & 2 x HDR200 & - & 4 x NDR200 & 1 x NDR200 & 1 x HDR100 & 1 x HDR100 \\
\textbf{Cooling} & Air & Air & DLC & DLC & DLC & DLC \\

\bottomrule
\end{tabular}
\end{table*}

Our benchmarking methodology involves testing the entire cluster every two weeks using ClusterBench. For different clusters and multiple repeated runs, we use a fixed set of pre-compiled binaries, reusing the same binary across repeated benchmarks to eliminate build-to-build variance; configuration settings (e.g., cluster and partition parameters) are adjusted according to the specific cluster and partition being tested. Each data point in the correlation plots, and each count in the distribution plots, represents one end-to-end benchmark run. We run each benchmark for at least 10 minutes to stress the components long enough. Benchmark performance numbers are obtained through reported performance numbers. For metric data points, we obtain a single data point by averaging the metrics collected during that benchmark run. Since the benchmarks run with stable workloads for a long enough time, the average converges to the expected value of the metric. The averaged data point is therefore a reliable summary of the benchmark run. In the normalized plots, the data has been normalized with respect to the range of observed metric values. 

In the distribution plots, the median value is reported. Along with the median value, the percentage fluctuation between the 5th and 95th percentiles of the distribution is also reported.

\subsection{Results}

The performance variation of dual-socket HPL runs is first examined on the largest CPU-only cluster at NHR@FAU. The \texttt{singlenode} partition on Fritz consists of 992 nodes with dual-socket Intel Ice Lake SP processors.

\begin{figure}[tbp]
    \centering
    \includegraphics[width=0.49\textwidth]{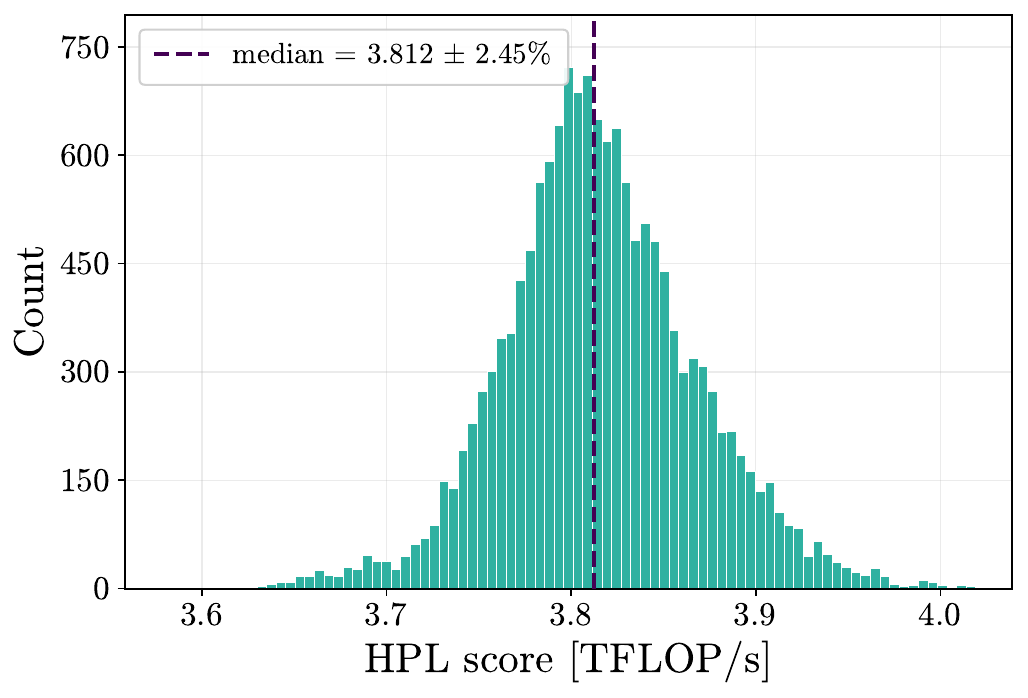}
    \caption{Per-node CPU HPL performance distribution across 992 nodes of \texttt{singlenode} partition on Fritz (Intel-optimized, May 2025–July 2026).}
    \label{fig:res1}
\end{figure}

Figure~\ref{fig:res1} shows a performance variation of up to 5\% around a mean of 3.8 TFlop/s. The peak throughput of the dual-socket Ice Lake chips in this partition is approximately 5.5 TFlop/s. This gap is explained by core frequency throttling: the base core frequency is 2.4 GHz, while the sustained frequency during HPL execution is 1.85 GHz. The distribution aggregates data from all nodes over one year, combining spatial and temporal variation. The absence of extreme outliers indicates that the majority of nodes are healthy.

\begin{figure}[tbp]
    \centering
    \includegraphics[width=0.49\textwidth]{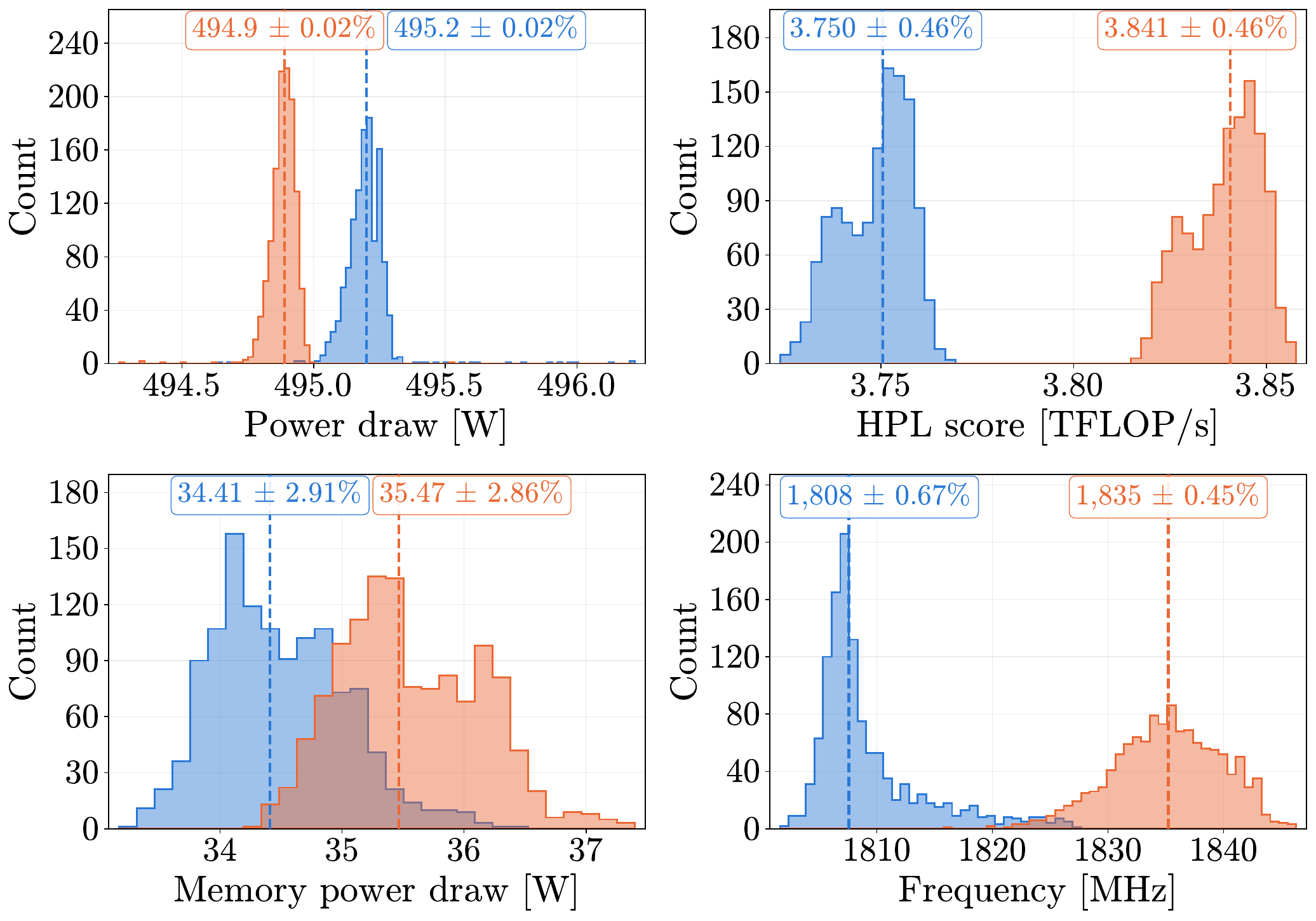}
    \caption{Distribution of CPU HPL performance and CPU metrics for nodes f0383 (blue) and f0384 (orange) of \texttt{singlenode} partition on Fritz (Intel-optimized, May 2025–July 2026).}
    \label{fig:res2}
\end{figure}

A closer examination of two individual nodes from the \texttt{singlenode} partition provides further insight. Figure~\ref{fig:res2} shows the per-node distribution of performance and related metrics. Each node exhibits a performance variation of less than 1\%, and each metric shows a distinct pattern of fluctuation. Node f0383 shows a slightly higher power draw and more frequency throttling than node f0384, which results in lower performance. These distributions indicate that each node has a characteristic operating signature.

\begin{figure}[tbp]
    \centering
    \includegraphics[width=0.49\textwidth]{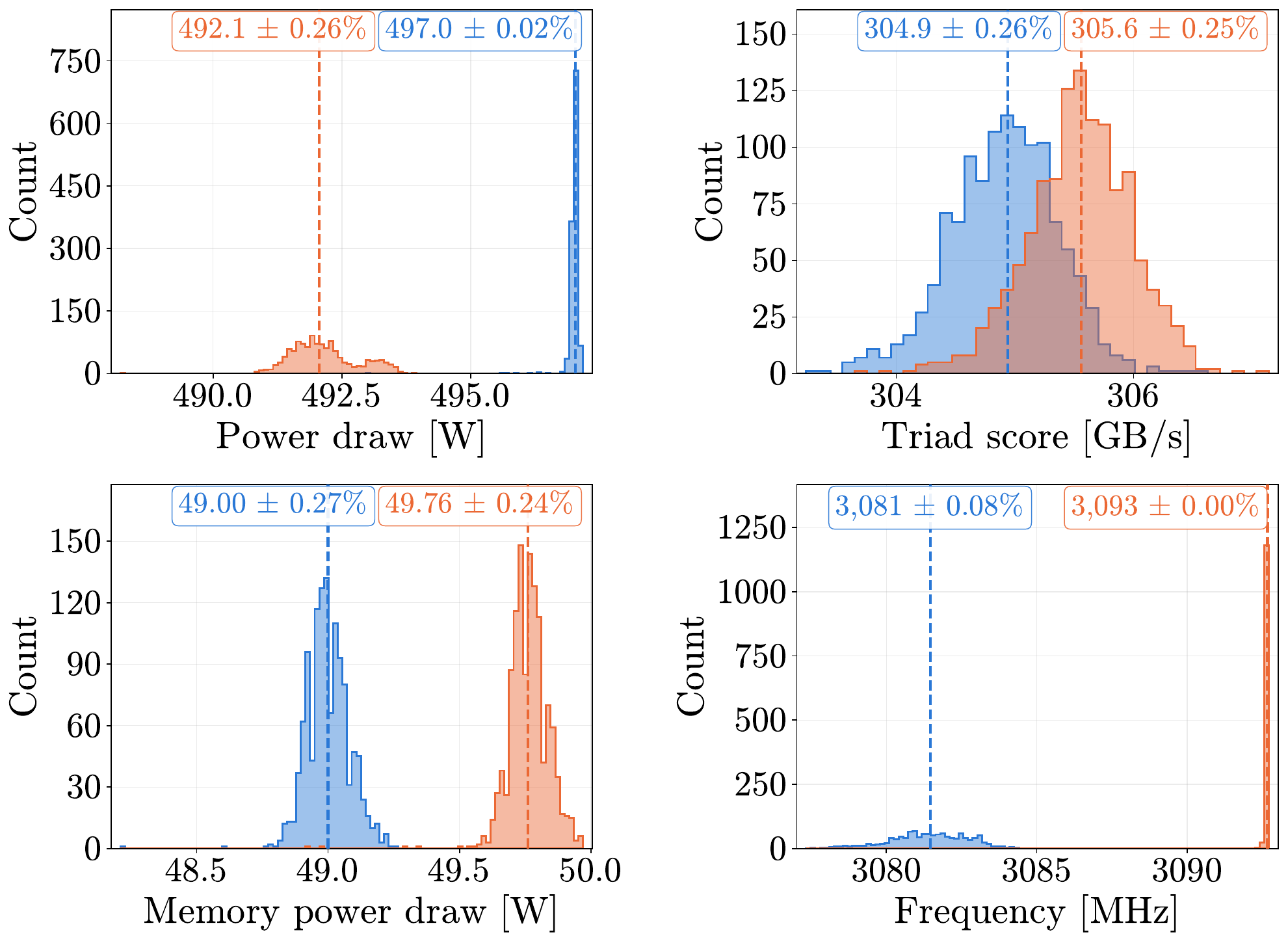}
    \caption{Distribution of CPU Schoenauer Triad kernel performance and CPU metrics for nodes f0383 (blue) and f0384 (orange) of \texttt{singlenode} partition on Fritz (May 2025–July 2026).}
    \label{fig:res3}
\end{figure}

Because HPL is compute bound and stresses the CPU, it is necessary to determine whether the same nodes behave differently under memory-bound stress. We use repeated calls to the Schoenauer Triad kernel (\texttt{a[i] = b[i] + c[i] * d[i]}) from TheBandwidthBenchmark, where \texttt{a}, \texttt{b}, \texttt{c}, and \texttt{d} are arrays. Such streaming kernels are sufficient to stress the memory interface. Figure~\ref{fig:res3} shows that the memory-bound code produces different characteristics on the same two nodes. The fluctuation in memory power draw is smaller than for HPL, although the spread of power draw remains similar; the shape of the distribution changes. The higher power draw on node f0383 leads to greater frequency fluctuation, whereas node f0384 draws less power, remains further from its TDP limit, and maintains a more stable frequency. For both compute-bound and memory-bound code, node f0383 underperforms node f0384 slightly, consistent with its higher power draw.

\begin{figure}[tbp]
    \centering
    \includegraphics[width=0.49\textwidth]{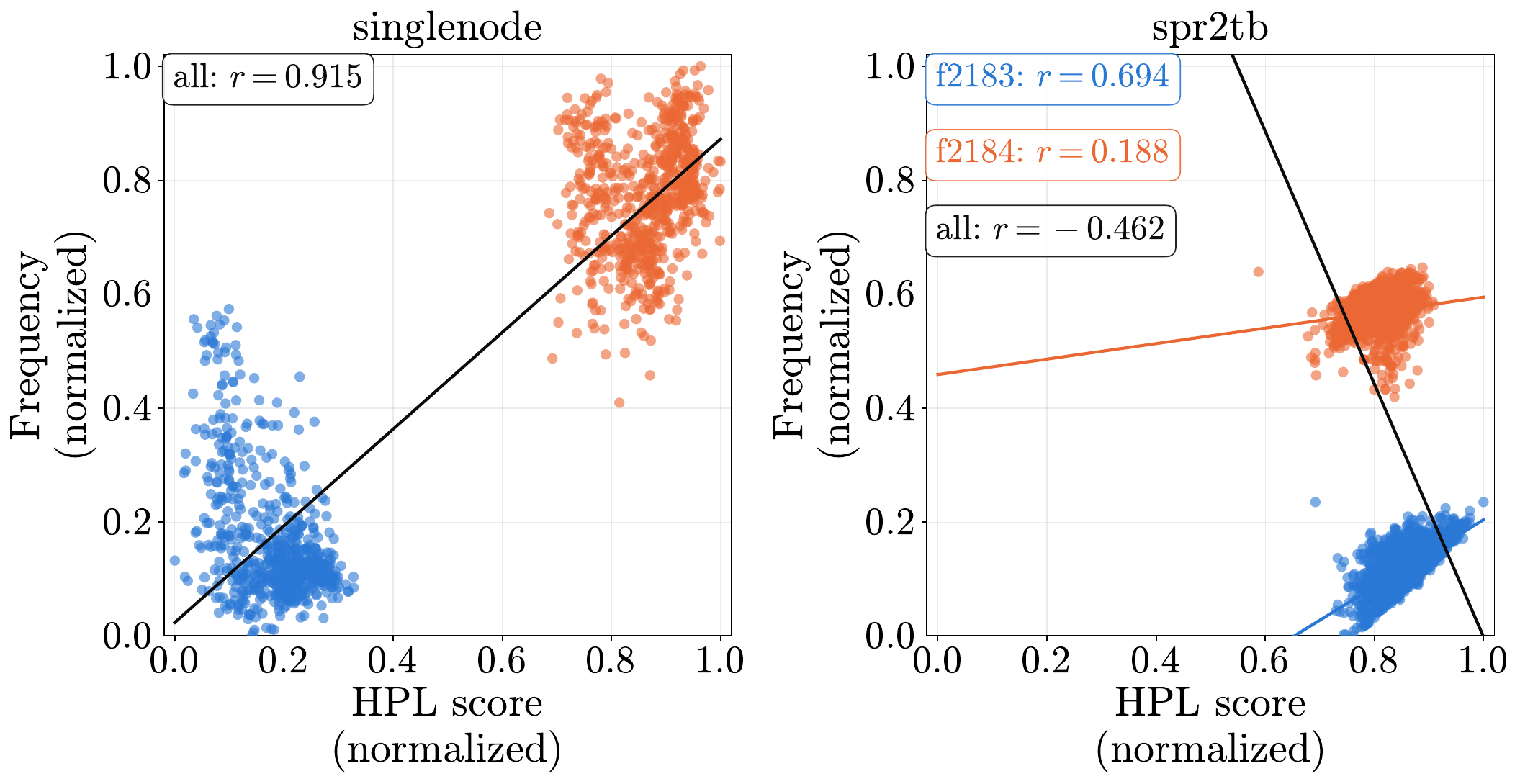}
    \caption{Correlation between CPU HPL performance and frequency for the \texttt{singlenode} partition (nodes f0383 (blue), f0384 (orange)) and \texttt{spr2tb} partition (nodes f2183 (blue), f2184 (orange)) on Fritz (Intel-optimized, May 2025–July 2026).}
    \label{fig:res4}
\end{figure}

These observations motivate an examination of correlations between metrics. In Figure~\ref{fig:res4}, each node occupies a distinct region in the correlation plot, forming a separate cluster. A positive correlation between frequency and performance is expected, since reduced frequency limits the throughput of compute-bound code. For the nodes in the \texttt{singlenode} partition, this relationship holds and confirms the expectation. However, data from two additional nodes in the \texttt{spr2tb} partition illustrate a risk in correlating metrics and performance at cluster level than at individual level. At the level of an individual node, frequency and performance remain positively correlated, as expected. When data from both nodes are combined across time, however, the aggregate correlation appears negative, which does not reflect the true relationship. This shows that correlation data at the individual node level might be different from the correlating data across the whole cluster at once. The outlier nodes in the cluster may sway the correlation to a certain extent. It also shows that each CPU has a characteristic operating signature, visible as a distinct cluster in the scatter plot. 

\begin{figure}[tbp]
    \centering
    \includegraphics[width=0.49\textwidth]{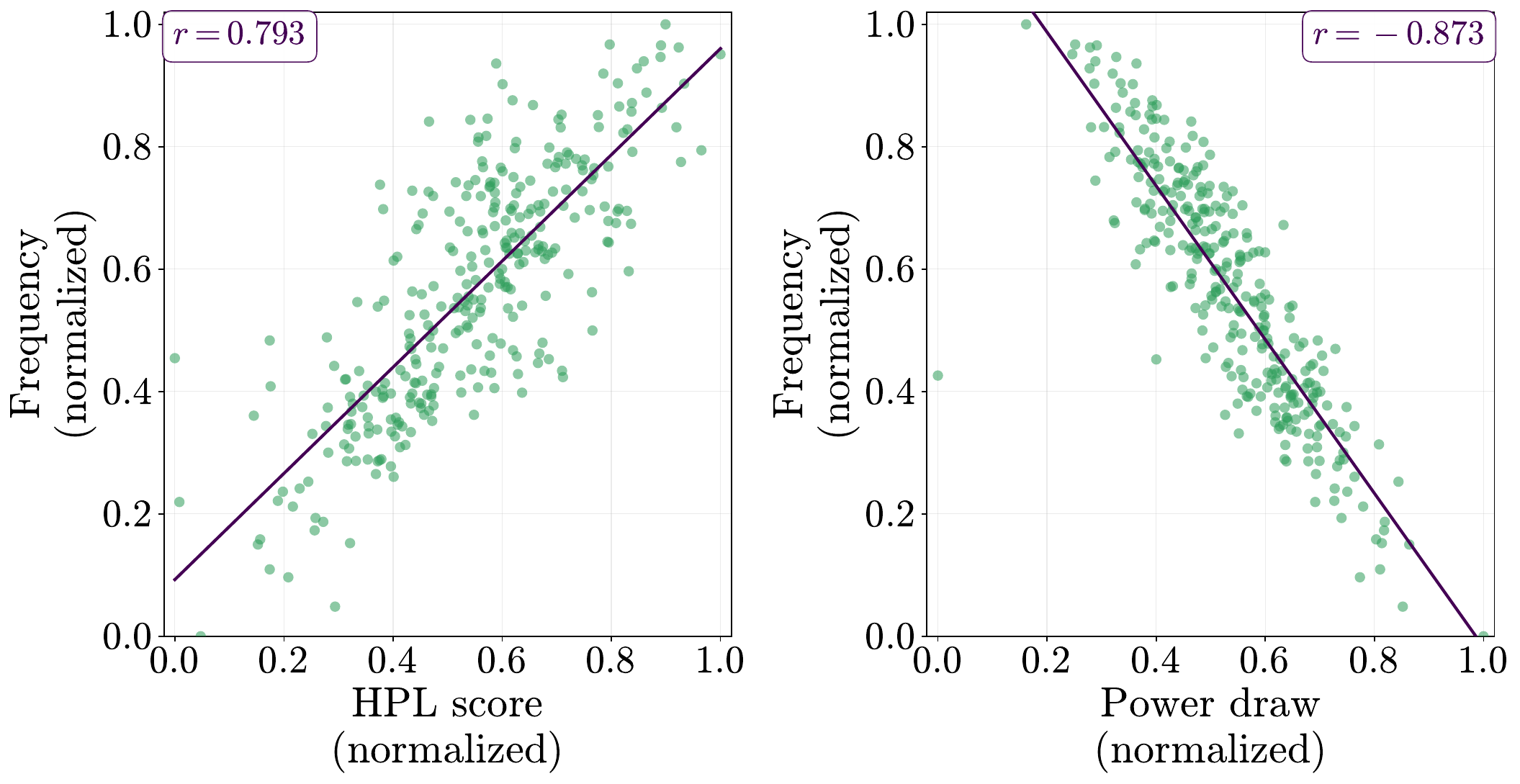}
    \caption{Correlation between per-node CPU HPL performance and CPU metrics across 300 nodes in the \texttt{cpu} partition on Helma (AMD-optimized, 2 July 2026).}
    \label{fig:res5}
\end{figure}

Dual-socket HPL runs were also performed on the AMD Turin Dense nodes (384 cores per node) in the \texttt{cpu} partition of the Helma cluster. Figure~\ref{fig:res5} presents a single-timestamp snapshot of the \texttt{cpu} partition, with one data point per node. Given the sample size is big enough, this confirms the expected positive correlation between frequency and performance. It also shows that higher power draw is associated with greater frequency throttling, producing a strong negative correlation between power draw and frequency for compute-bound code. Because each node forms a distinct cluster in the correlation plot, restricting the analysis to one point in time per node provides a more reliable method for computing cluster-wide correlations between metrics and performance. 

\begin{figure}[tbp]
    \centering
    \includegraphics[width=0.49\textwidth]{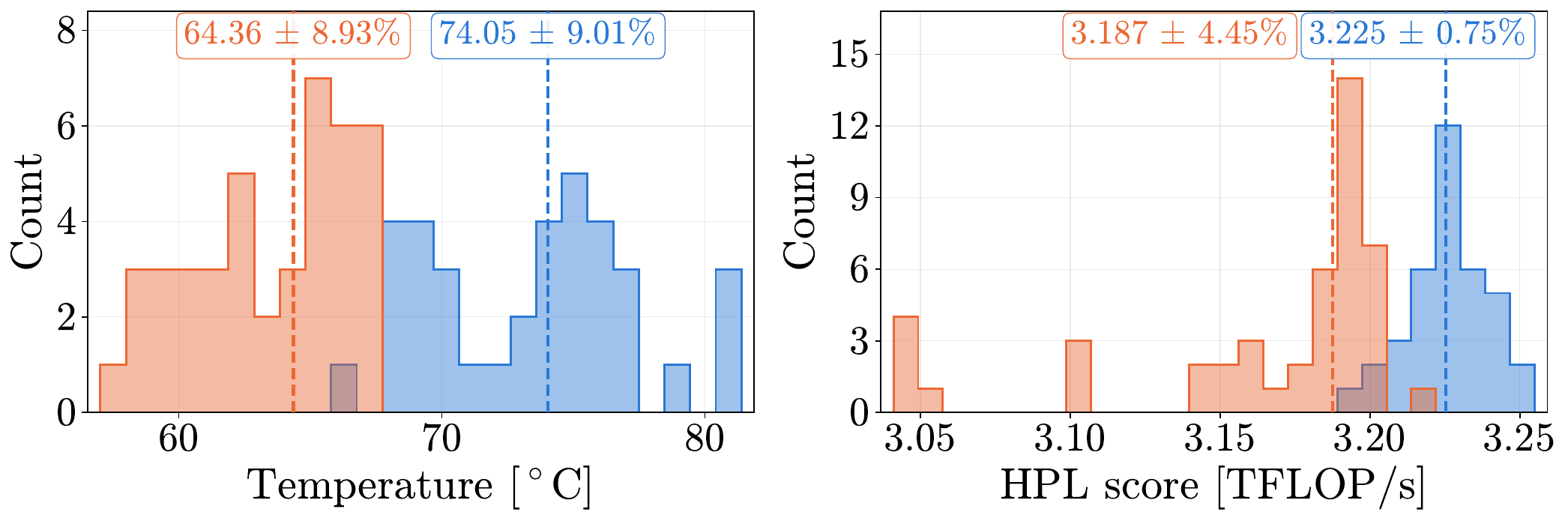}
    \caption{Distribution of per-node CPU HPL performance and CPU temperature across 64 nodes in the \texttt{a40} and \texttt{a100} partition of Alex (AMD-optimized, 31 Jan 2026).}
    \label{fig:res6}
\end{figure}

The Alex cluster contains two partitions, \texttt{a40} and \texttt{a100}, named for their NVIDIA A40 and NVIDIA A100 GPUs. The cluster is air-cooled. We evaluated the host CPUs of both partitions under stress testing. Figure~\ref{fig:res6} shows that, despite using identical CPU models, the CPUs in the \texttt{a100} partition run approximately 10 degrees hotter than the CPUs in the \texttt{a40} partition. Higher temperature would be expected to increase throttling and reduce performance; instead, the CPUs in the \texttt{a100} partition perform slightly better. This result is unexpected, and the underlying cause remains unclear. 

\begin{figure}[tbp]
    \centering
    \includegraphics[width=0.49\textwidth]{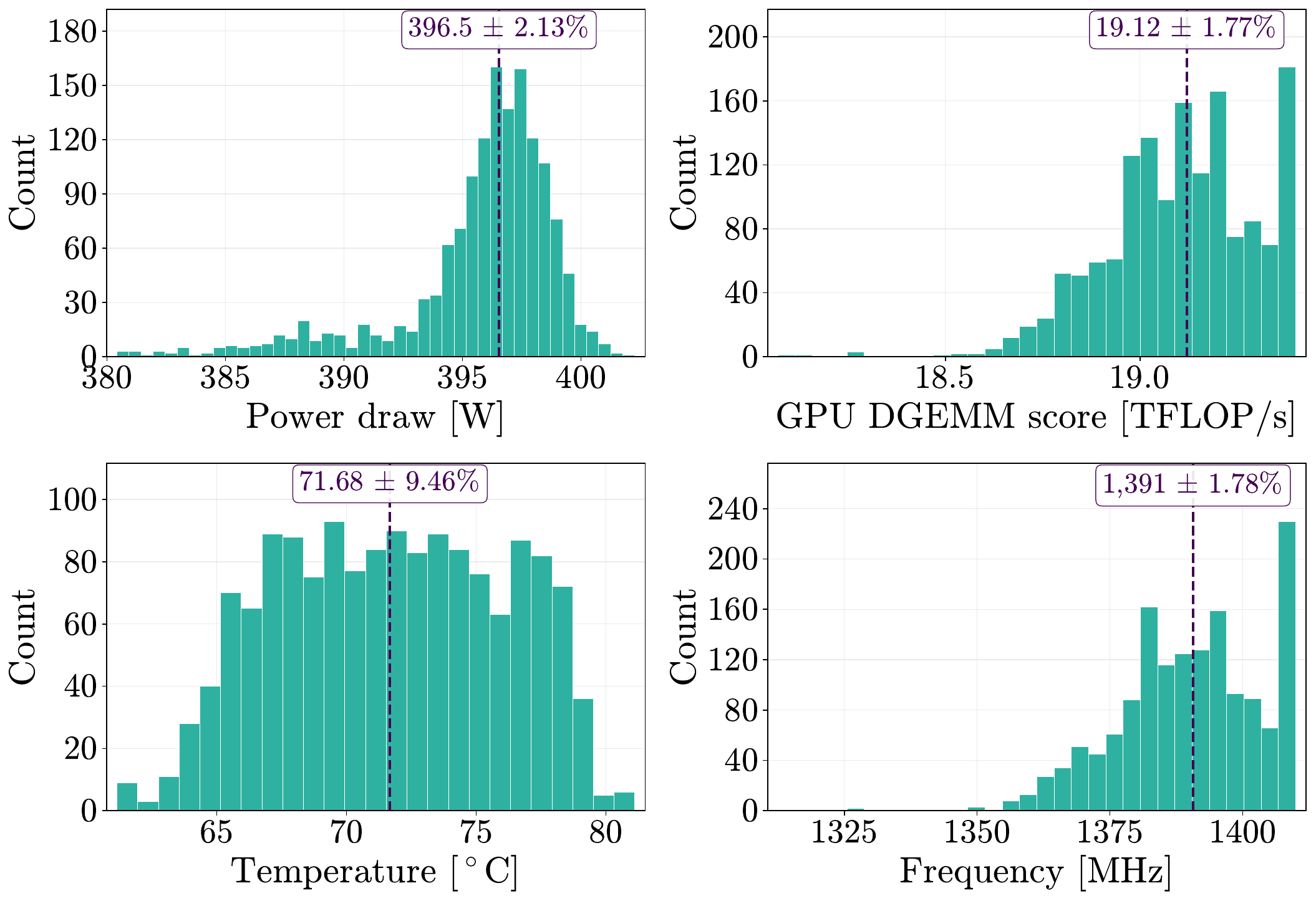}
    \caption{Distribution of per-GPU DGEMM performance and GPU metrics for all 304 GPUs (38 nodes) of the \texttt{a100} partition on Alex (CuBLAS-optimized, Nov 2025–July 2026).}
    \label{fig:res7}
\end{figure}

We also examined compute-bound and memory-bound codes on GPUs, which show different performance variability depending on code type. Figure~\ref{fig:res7} shows the performance variability of the compute-bound GEMM kernel on the A100 GPUs. An identical DGEMM binary was submitted to each individual GPU within the node. Performance varies by less than 5\% across space and time. The distribution of SM frequency closely matches the distribution of DGEMM performance, indicating a strong correlation between the two. GPU temperature shows the highest variability of the measured metrics, consistent with the air-cooled design of these nodes. Figure~\ref{fig:res9} examines the correlation between performance, frequency, and temperature in more detail. Before turning to these correlation plots, however, the distributional analysis of memory-bound code on GPUs merits further discussion.

\begin{figure}[tbp]
    \centering
    \includegraphics[width=0.49\textwidth]{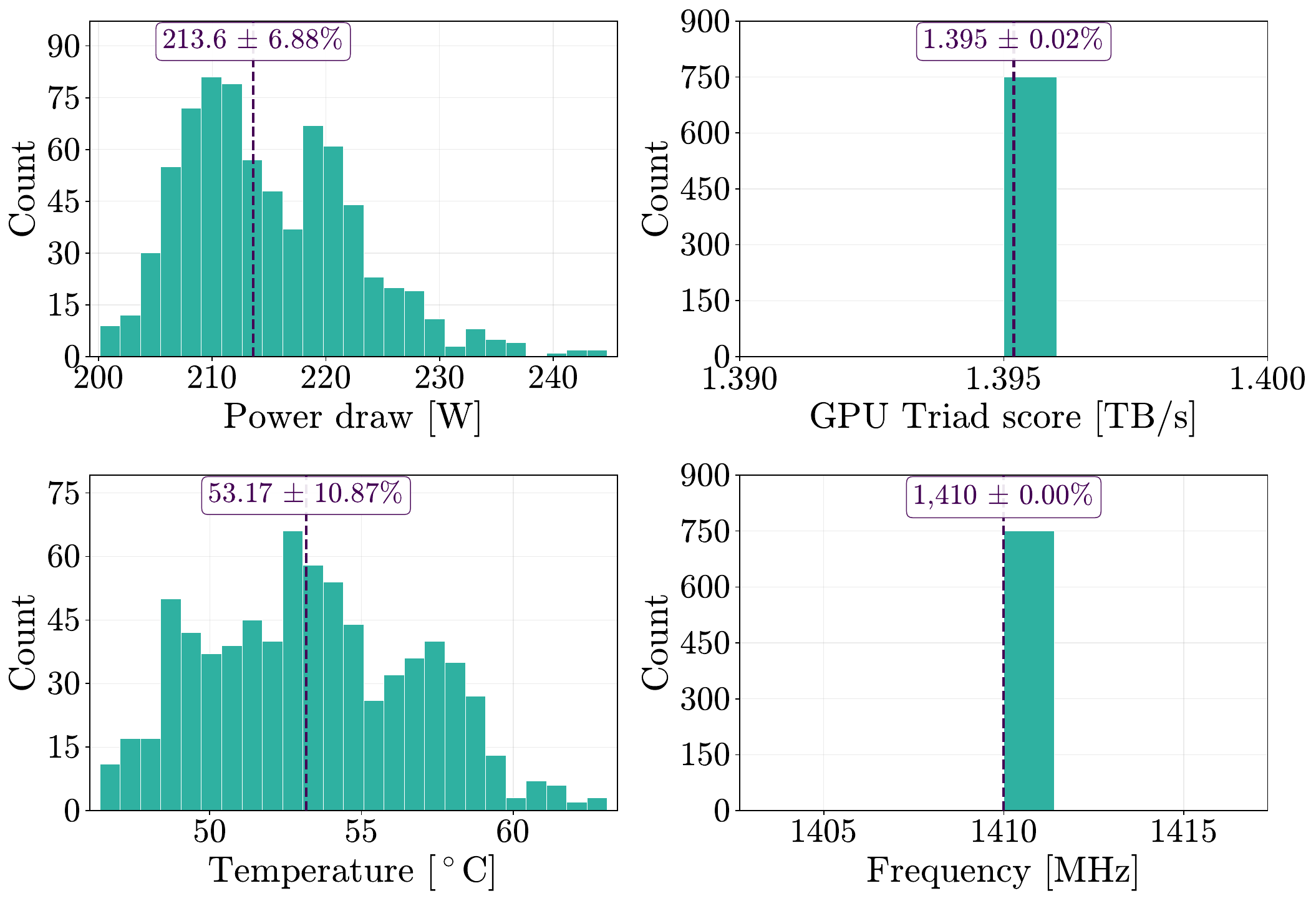}
    \caption{Distribution of per-GPU Schoenauer Triad performance and GPU metrics for all  160 GPUs (20 nodes) of the \texttt{a100} partition on Alex (Nov 2025–July 2026).}
    \label{fig:res8}
\end{figure}

To stress the memory hierarchy and components of the GPU, the Schoenauer Triad kernel from TheBandwidthBenchmark is again used. Figure~\ref{fig:res8} shows stable Schoenauer Triad performance for the GPUs. This performance stability, observed across all 160 GPUs (20 nodes) over time, suggests that these GPUs retain substantial power headroom. Although the distribution of GPU power draw shows high variation, the GPUs do not reach their TDP limit. Consequently, SM frequency does not throttle, and Schoenauer Triad performance remains stable. GPU temperature again shows high variability, consistent with the DGEMM results. Such stable Schoenauer Triad performance with less than 0.05\% variability was also observed on other GPU types in the tested clusters, including the NVIDIA A40, H100, and H200.

\begin{figure}[tbp]
    \centering
    \includegraphics[width=0.49\textwidth]{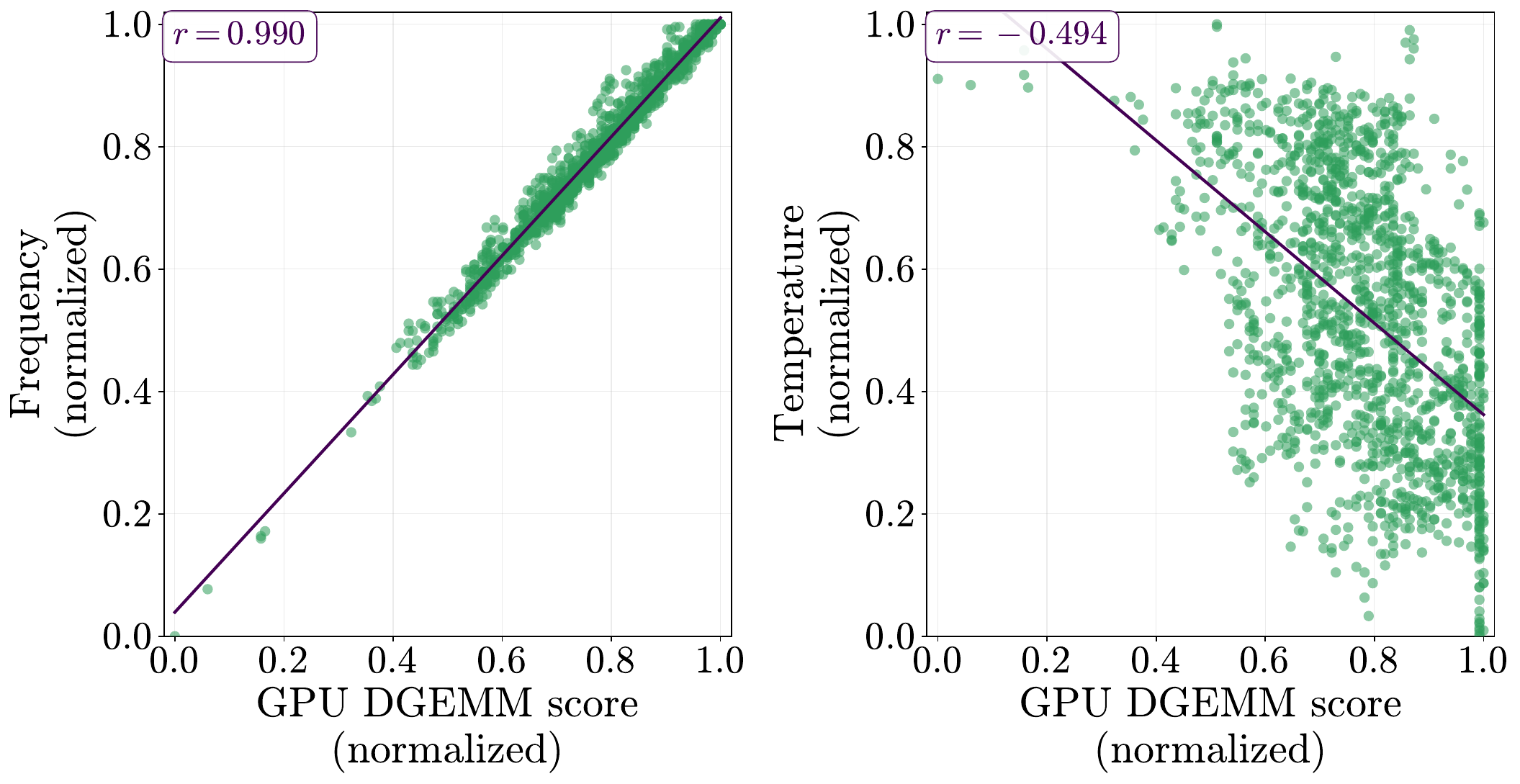}
    \caption{Correlation between per-GPU DGEMM performance and GPU metrics across all 304 GPUs (38 nodes) in the \texttt{a100} partition on Alex (CuBLAS-optimized, Nov 2025–July 2026).}
    \label{fig:res9}
\end{figure}

Correlation plots allow us to test two assumptions drawn from the GPU DGEMM distribution plots above: the dependence of the DGEMM performance on SM frequency and GPU temperature. Figure~\ref{fig:res9} confirms both. SM frequency correlates strongly and positively with performance, consistent with expectation. GPU temperature correlates negatively with performance, a relationship that is more complex and cluster dependent. Because the \texttt{a100} partition is air cooled, thermal throttling is a contributing factor: higher temperature increases SM frequency throttling, which in turn reduces the DGEMM performance. This relationship does not hold for the GPUs in the \texttt{h100} partition.

\begin{figure}[tbp]
    \centering
    \includegraphics[width=0.49\textwidth]{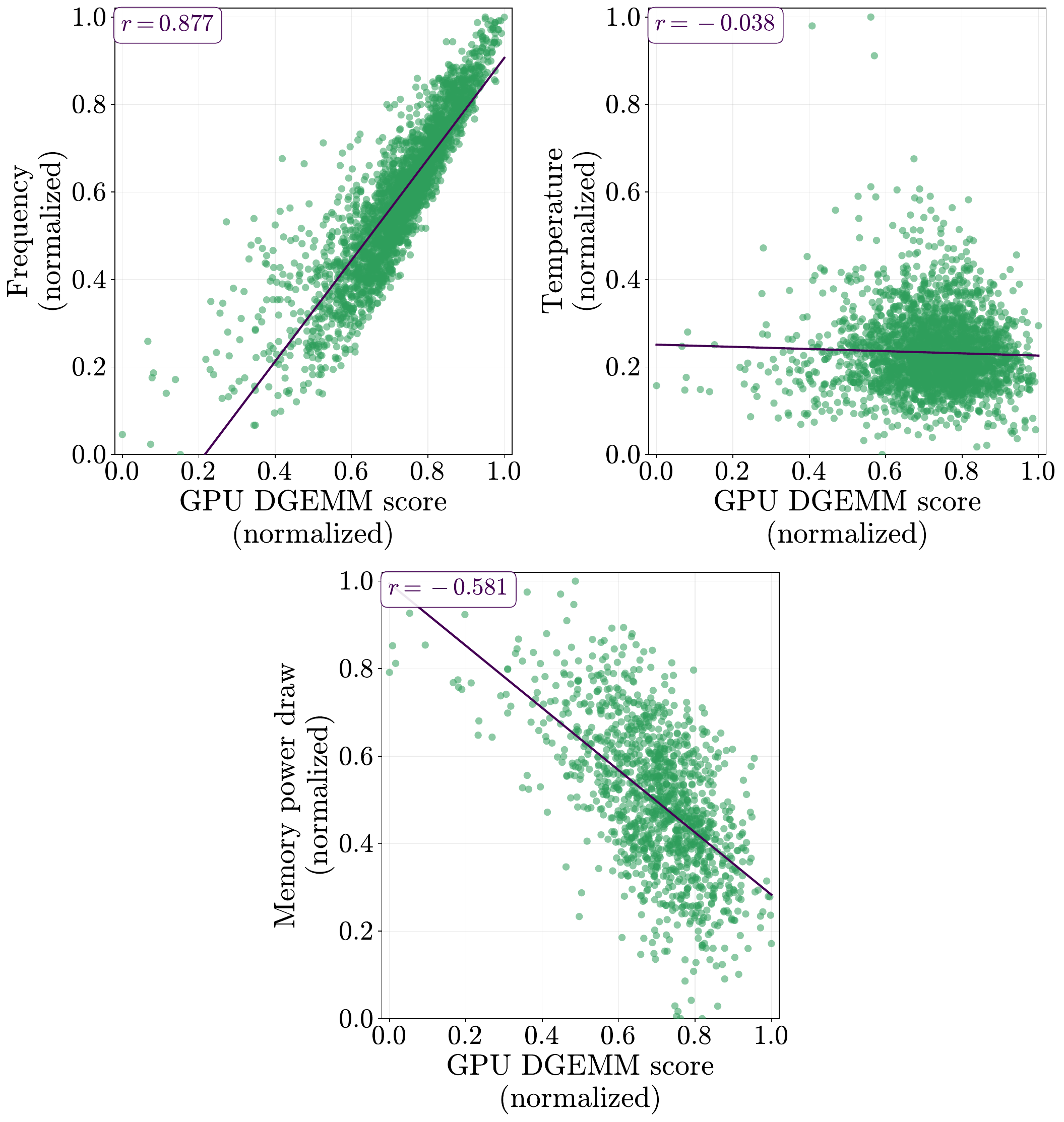}
    \caption{Correlation between per-GPU DGEMM performance and GPU metrics across all 384 GPUs (96 nodes) in the \texttt{h100} partition on Helma (CuBLAS-optimized, Jan 2026–July 2026).}
    \label{fig:res10}
\end{figure}

The NVIDIA H100 GPUs in the Helma cluster are water cooled and show a different relationship between GPU temperature and DGEMM performance. Figure~\ref{fig:res10} shows no correlation between GPU temperature and DGEMM performance in this case, even though these GPUs also reach their TDP limit and experience SM frequency throttling. The median for the H100 GPU temperature was \qty{53 +- 2}{\degreeCelsius}. The strong correlation between SM frequency and benchmark performance observed previously also holds for the H100 GPUs. The correlation between memory power draw and DGEMM score reveals an additional pattern. Because GPU memory power draw is included in the total GPU power budget, an increase in GPU memory power draw reduces the power available to other components, lowering SM frequency and, consequently, performance~\cite{arch_tradeoff}. This produces the observed negative correlation between DGEMM performance and GPU memory power draw. This relationship does not hold for codes that do not reach TDP and therefore do not experience SM frequency throttling; in those cases, GPU memory power draw and performance show no correlation.

These results raise several questions for future investigation. How frequently should component-level health checks be performed: daily, weekly, or monthly? How does the observed distribution evolve over time, whether across years or over the operational lifetime of a cluster? What machine learning or statistical methods are best suited to capture performance fluctuations or degradation? A complete treatment of these questions requires additional time and data and is left for future work.

\section{Limitations and Future Work for ClusterBench}
\label{sec:limitations}

ClusterBench was designed for a narrow purpose: repeated, component-level measurement of every node in a production system, whether for cluster validation or ongoing health monitoring. Several capabilities that a general-purpose benchmarking framework would offer are therefore absent by design in ClusterBench rather than by oversight. This section states those boundaries, the reasoning behind them, and the work planned beyond them.

\subsection{Compilation phase}
\label{sec:compilation_phase}
We have deliberately kept the compilation phase out of this framework. Since the target is to perform cluster-wide component-level testing, the burden to provide correct binaries falls onto the user. The user needs to provide the binary and provide the correct path to it. Compilation phase implementation adds lots of complexity, for which this framework was never designed. Production environments do not change rapidly, so the dependent modules or compilers would be present for a single binary. Recompiling the benchmarks for every run is unnecessary when the health check is executed weekly. It adds build overhead to each cycle. Compilation behavior is also not the target of the measurement: the goal is to stress the components, not to characterize the build. The binaries are therefore built once and reused across runs. MachineState allows capturing the software environment to a certain extent, but documentation of the configuration management for the binaries, e.g., benchmark version, build environment, dependencies (MPI, CUDA), etc., is not captured. If required, there are existing frameworks that can be coupled with ClusterBench to function as an end-to-end regression testing framework.

\subsection{Parameter space exploration}
ClusterBench does not sweep input parameters or ranges. A component-level health check answers a narrower question, namely whether a node behaves like its peers under one fixed configuration, and holding that configuration constant is what makes results comparable across nodes and across time. A sweep would multiply the node count by the size of the parameter space, which is difficult to justify against a production allocation for routine monitoring. However, the capability is useful for characterization work and its addition is planned.

\subsection{Scheduler support}
Slurm is the most widely used scheduler in HPC environments today, and ClusterBench currently supports only Slurm. The scheduler-facing logic is limited to three tasks: submission, state polling, and accounting queries. Because this interface is narrow, support for other schedulers can be added later as separate backends without changing the core logic of the framework. PBS is an older scheduler and is used less often now. Flux is newer and not yet widely used in production systems; support for either can be added if a clear need arises. HTCondor comes from the AI community and is mature, so it may be worth supporting if it becomes more common in HPC settings.

\section{Conclusion}
\label{sec:conclusion}

This paper presented ClusterBench, a framework for cluster-wide continuous benchmarking and health monitoring of production systems. The framework treats the cluster, rather than a single node, as the object under test. A test definition is set up once for a cluster or a partition, and ClusterBench derives from that definition the submission of identical benchmarks to every node, to node pairs for interconnect measurements, and to each accelerator within a node. Each result is stored together with the node, the timestamp, and the software environment, and every benchmark is wrapped in a metric collector that records power draw, core frequency, and temperature over the same interval. Results therefore accumulate along two axes, across the nodes and across their lifetime.

Using the accompanying component-level benchmark suite, we observed the NHR@FAU clusters Fritz, Alex, and Helma over more than one year. Performance variation within a single component stays below 1\%, while variation across component specimens of the same type reaches up to 5\% on nodes that are identical by specification. The distributions of performance and of the accompanying metrics are narrow and reproducible per node, which indicates that each component has a characteristic operating pattern rather than a single nominal value. The absence of extreme outliers in these distributions is itself the result a data center needs from an acceptance or health check.

The correlation between performance and collected metrics strongly depends on how the data is aggregated. Within a single node, frequency and performance are positively correlated, as expected for compute-bound code. Once measurements from multiple nodes are combined across time, however, this correlation can invert, no longer reflecting the underlying per-node relationship. A snapshot taken at one point in time, with one measurement per node, is therefore the more reliable basis for such an analysis. The relationship between temperature and performance further depends on the node design: it is clearly negative on the air-cooled A100 nodes of Alex and absent on the water-cooled H100 nodes of Helma. Memory-bound code on GPUs behaves differently again compared to CPUs, remaining stable where the TDP limit is not hit and frequency does not throttle.

Comparisons across runs are therefore easier to interpret when the observed performance and metric spread for the given cluster is known, and continuous measurement is one way to obtain it. The dataset accumulated in this way also serves a second purpose beyond validation, as a long-term record for the study of hardware variability and degradation. Future work will extend the framework along the directions outlined in Section~\ref{sec:limitations} and use the growing dataset to determine suitable intervals for component-level health checks and advanced statistical/machine learning methods for detecting gradual degradation.

%\section*{Authors and Affiliations}

\section*{Artifact Evaluation}
The ClusterBench software, all the statistical data from the Results section, the compiler flags used to build and the exact benchmark parameters to run the benchmarks in this paper are all available through Zenodo at \url{https://doi.org/10.5281/zenodo.21870554}. For verifying the statistical data, a simple dashboard and the duckdb database object is also shipped in the given DOI repository. Detailed information can be further provided upon request. 

\section*{Acknowledgment}

The authors gratefully acknowledge the scientific support and HPC resources provided by the Erlangen National HPC Center (NHR@FAU) of the Friedrich-Alexander-Universität Erlangen-Nürnberg (FAU) under the BayernKI project v111dc. BayernKI funding is provided by Bavarian state authorities. The authors gratefully acknowledge the scientific support and HPC resources provided by NHR@FAU of the FAU under the NHR project b104dc. NHR funding is provided by federal and Bavarian state authorities. NHR@FAU hardware is partially funded by the German Research Foundation (DFG) -- 440719683.

\bibliography{ref}
\bibliographystyle{ieeetr}

\end{document}